# *In situ* tensile rupture test of medial arterial tissue in X-ray micro-tomography

Clémentine Helfenstein-Didier[1], Damien Taïnoff[2], Julien Viville[2], Jérôme Adrien[2], Éric Maire[2], Pierre Badel[1]

[1] Univ Lyon, IMT Mines Saint-Etienne, Centre CIS, INSERM, SainBioSE, F - 42023 Saint-Etienne FRANCE

[2] Université de Lyon, INSA-Lyon, MATEIS CNRS UMR5510, Villeurbanne, France

**\*Corresponding Author:**    Pierre Badel, PhD., badel@emse.fr

**Corresponding Author's address:**

Centre Ingénierie et Saanté

158 Cours Fauriel

42 023 SAINT-ETIENNE Cedex 2

Tel: +33 4 77 42 02 60

**Manuscript Region of Origin:**    France

**Word count (Introduction through Conclusion): 3211**

# Abstract

Detailed characterization of damage and rupture mechanics of arteries is one the current challenges in vascular biomechanics, which requires developing suitable experimental approaches. This paper introduces an approach using *in situ* tensile tests in an X-ray micro-tomography setup to observe mechanisms of damage initiation and progression in medial layers of porcine aortic samples. The technique requires the use of sodium polytungstate as a contrast agent, of which the conditions for use are detailed in this paper. Immersion of the samples during 24 hours in a 15 g.L$^{-1}$ concentrated solution provided the best compromise for viewing musculo-elastic units in this tissue. The process of damage initiation, delamination and rupture of medial tissue under tensile loading was observed and can be described as an elementary process repeating several times until complete failure. This elementary process initiates with a sudden mode I fracture of a group of musculo-elastic units, followed by an elastic recoil of these units, causing mode II separation of these, hence a delamination plane. The presented experimental approach constitutes a basis for observation of other constituents, or for investigations on other tissues and damage mechanisms.

# 1. Introduction

Detailed characterization of damage and rupture properties of arteries is one the current challenges in vascular biomechanics. Several approaches to characterize rupture and/or dissection properties of vascular tissues have been addressed (see for instance, the review (Tong et al., 2016) and references herein). A large proportion of this work was based on measured properties at the macroscopic scale that could serve the global understanding of the phenomenon and the development of constitutive models (Gasser et al., 2006). The use of histology helped in providing hints about the local effects of damage progression through the tissue. Towards a more complete description and understanding of arterial mechanics and damage, studies involving simultaneous mechanical loading and observations at the scales of the microstructure are necessary. Such studies are still scarce in the literature.

Recently, several groups used multiphoton confocal microscopy (MPCM) to image the internal structure of arterial tissue (e.g. Rezakhaniha et al., 2012; Wang et al., 2013; Robertson et al., 2015). In these studies, the authors most often focused on collagen structures which are particularly (Robertson et al. 2015)suitable to be imaged using this type of microscope. These pioneering studies were the first to show the evolving geometry of collagen inside the arterial tissue under uniaxial and biaxial loading tests. Though this approach is very interesting and promising to provide deeper insight in the micro-scale mechanisms driving the mechanical response of such tissue, it turns out to be very difficult to setup, especially when combining it with mechanical tests. Its use in observing damage mechanisms still represents a challenge at the moment.

Among other alternative non-destructive techniques, X-ray micro-tomography (XRCT) is probably one of the most accessible. 3D observations of biological tissues using this technique have already been proposed in the literature (Metscher et al., 2009a, 2009b, 2011; Johnson et al., 2006; Wirkner and Prendini, 2007). However, these studies were performed at larger scales. Other groups (Pauwels et al., 2013; Aslanidi et al., 2013a, 2013b; Jeffery et al., 2011; Mizutani et al., 2012; Walton et al., 2015) used high-Z element staining to investigate the structure of soft tissues at lower scales, for instance. However, none of them performed *in situ* tests, because, in Walton et al. (2015) for instance, specimens had to be paraffin-embedded and acquisition time was between 5 and 16 hours for one sample preventing from performing such tests.

The scale that can be imaged using this technique is slightly larger than that of MPCM, but it offers an intermediate level that may be relevant for the study of arterial tissue, and for instance to investigate mechanisms like the propagation of dissection. In particular, it allows observing the whole thickness of human-size large arteries with a high resolution. This compromise is relevant to keep an overview of the sample while studying details of its structure. Collagen structures of vascular wall were recently investigated using XRCT with various staining techniques of the tissues to yield appropriate contrast (Nierenberger et al., 2015). *In situ* mechanical testing in an XRCT equipment is now widely spread (Buffiere et al., 2010) in materials science. For soft tissues, such *in situ* testing is likely to be less complex than in an MPCM setup which requires perfect stability of the sample. For these reasons, XRCT which is a more and more accessible technique in research laboratories seems to be a relevant tool to study meso- or even micro-scale mechanics of arterial tissue.

In this context, the present study was carried out on a conventional XRCT equipment and aimed at (i) studying the feasibility of combining *in situ* mechanical testing up to rupture and structural observations of arterial tissues and (ii) providing clues on rupture mechanisms that could be observed. First, this study evaluated the technical conditions required to perform such observations in the medial layer of porcine aortic tissue, the layer where dissection damage initiates and propagates. Then, *in situ* tensile tests, up to failure of the tissue, were carried out representing a possible *in vitro* model of dissection propagation under tensile loading.

## 2. Materials and methods

### 2.1 Sample preparation

Six porcine hearts with aortas were obtained from a local slaughterhouse few hours after death.

After careful dissection of the aorta, 48 dog-bone shaped thoracic aortic samples were cut with a manual press and a custom punch that ensured repeatability of the cuts. 15 of them were cut along the circumferential direction and 33 along the longitudinal direction. The outer layer of each sample, the adventitia, was carefully removed in this study focusing on the media only. Histological investigation on one preliminary test sample confirmed that this removal step was properly executed. All samples were then conserved, immersed, in a refrigerator at 5°C until testing (for typically 24 or 48 hours depending on the staining protocol, see below). The internal layer, the intima (one very thin layer of cells), could not be distinguished in these samples and was kept according to the common assumption that it has a negligible mechanical role.

### 2.2 X-ray microtomography

#### 2.2.1 X-ray microtomography setup

XRCT was used to image the samples with or without applied load. The principle of XRCT is explained in details in (Baruchel et al., 2000). This technique is analogous to the medical scanner, with much lower dose, and allows reconstructing non-destructively the internal structure of an opaque material from a set of X-ray attenuation radiographs. The reconstruction involves a computed step and the final image, in the case of laboratory based XRCT, is a 3D map of the local X-ray attenuation coefficient.

The tomograph used in this study is a commercial model which includes a nanofocus transmission X-ray tube (W target). The size of the focus (and thus the highest achievable resolution) is tunable from 1 to 5 µm. The detector used is an amorphous silicon flat panel composed of 1536 x 1920 square pixels with a lateral size of 127 µm. The setup exhibiting cone beam geometry, the sample is simply placed at different distances from the source to obtain different magnification levels. The lowest voxel size achieved in this study is 4 µm because of the tradeoff between magnification and maximum specimen dimension.

Soft tissues are not as stable as more commonly used hard samples. The samples are thus likely to move during scanning and this motion should be minimized as it would result in a blurry reconstructed image. Fast acquisition is thus mandatory especially in the case of *in situ* tensile experiments. With the purpose of minimizing exposure time, scanning times of 5 to 9 minutes for a voxel size of 7 µm and about 15 min for a voxel size of 4 µm were obtained, with the beam being operated at 280 µA and 80 kV. The exposure time was about 200 ms for each projection.

The uncertainty of the dimensions measured by XRCT is influenced by many parameters (mainly quality of the image acquisition and processing) and was estimated to ±1 voxel.

### 2.2.2 Absorption contrast in these soft tissues

As confirmed by a preliminary scanning test, arterial tissue does not naturally exhibit sufficient absorption contrast to obtain satisfying images. For this reason, it was necessary to use a contrast agent that binds with some components of the material (note that phase contrast is a possible alternative that requires specific equipment which is much less common than absorption contrast equipment). Solutions of diluted sodium polytungstate (SPT) (Nakashima, 2013; Nakashima and Nakano, 2014) were used to increase the absorption coefficient of elastin sheets of the medial layer in this arterial tissue (the medial layer is mainly composed of a superposition of elastin sheets and smooth muscle cell layers forming the so-called musculo-elastic units (Humphrey, 2002). A complete study of this staining technique and its possible influence on the mechanics of the problem is presented below in section 2.4.

## 2.3 *In situ* mechanical testing

### 2.3.1 Tensile testing

A custom tensile testing setup (described in Buffiere et al. (2010) and satisfying the constraints of use in an XRCT equipment) connected to a 200 N load cell was positioned in the XRCT. Dog-bone shaped samples (6 mm long and 4 mm wide region of interest) were clamped at each end between two flat surfaces with sandpaper glued on it to improve the adhesion and the positioning. All *in situ* tensile tests were performed up to failure, at 0.02 mm/s while recording tensile force and displacement.

Nominal strain was calculated as $\Delta l/l_0$ with $\Delta l$ the cumulated displacement and $l_0$ = 6 mm the initial dimension of the sample. Nominal stress ($F/S_0$) was calculated using the average initial section measured on the 3D scan performed at no load ($S_0$) and the recorded tensile force (F). Though it is possible to accurately calculate the Cauchy stress in principle, it was not calculated because too few 3D scans were performed during the entire *in situ* test (to avoid prolonged X-ray exposure time and possible drying effects).

The *in situ* testing protocol was the following. First, an initial high resolution scan (4 µm) at no load was performed. Then, continuous radiograph acquisition was turned on, and a larger field of view was setup (hence a lower resolution, 20 µm) during the *in situ* loading. The aim was to keep a continuous observation, though not 3D, of the sample during the test. For these radiographic observations, the tangent plane of the sample was oriented parallel to the X–ray beam to obtain a lateral view of the layered structure. At partial rupture states (detected from the continuous force record and visual observations), the tensile test was stopped and the sample partially unloaded to prevent motion and/or damage propagation. 3D scans of the failure area could then be acquired at high resolution (4 µm). These 3D images could be used for further analysis of damage progression through the sample. The sample was then subsequently loaded with continuous radiograph acquisition up to the complete failure.

Six samples were tested *in situ*, see section 3 and Table 1 for details on these samples.

### 2.3.2 Histological investigation

Two samples were used for histological observations in order to complete the XRCT observations and confirm or correct the conclusions drawn regarding damage propagation mechanisms. The samples were fixed in alcohol solution (80%), embedded in MMA, sectioned at ~7µm and stained with orcein.

## 2.4 Protocol adjustment

Prior to performing the *in situ* tensile tests presented in section 2.3, most of the samples were used to adjust and evaluate the staining technique presented in this paper. Indeed, a contrast agent was used in order to obtain satisfying contrast and enable relevant observations. The conditions for use of this contrast agent and its possible effects on mechanical data had to be evaluated. For this, studies of penetration and mechanical effect of the contrast agent were performed. They are presented in sections 3.1 and 3.2.

### 2.4.1 Contrast agent penetration

24 samples were used to study the penetration of SPT during a concentration-controlled and time-controlled immersion of the samples. In this study, no mechanical testing was performed during XRCT imaging.

Three solutions with different concentrations of SPT were tested ($c_1$ = 5, $c_2$ = 10 and $c_3$ = 15 g.L$^{-1}$). For each solution, six samples were immersed. A commonly used saline solution (for control) and two high-concentration solutions of SPT ($c_4$ = 20 and $c_6$ = 30 g.L$^{-1}$) were also tested with only two samples for each. 3D scans were performed for each sample before immersion, and after immersion for 24 h (for all solutions) and 48 h (only for saline, $c_1$, $c_2$, $c_3$ solutions). During immersion, samples were stored at 5°C.

To avoid site-related variability, attention was paid to cutting the samples from the same aortic segment (ascending or descending aorta) when immersed in the same solution type.

The analysis of the obtained images was based on the measurement of the stained cross-sectional area in the samples. Penetration of SPT was quantified by calculating (i) the global stained area fraction and (ii) the radial and tangent stained depth based on the measurements of x1, x2, y1, y2 (see Figure 1), with tangent stained depth = (x1 + x2) / 2 and radial stained depth = (y1 + y2) / 2. Radial and tangent directions are respectively perpendicular and parallel to the layers of the medial microstructure. Statistical Wilcoxon (when n = 6) and Mann-Whitney tests were performed to study the penetration of the agent in the different conditions tested.

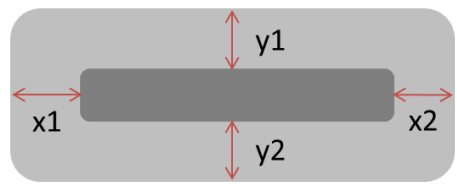

**Figure 1: Schematic view of a partially stained cross-section: measurements for normalized stained depth determination**

### 2.4.2 Mechanical effects of contrast agent

The effects of SPT on the mechanical tensile response of medial tissue were evaluated through tensile tests performed up to failure (no XRCT imaging). 18 samples were used to investigate these possible effects (see Table 1). Three groups of six samples were tested, with attention to cut them from the same aortic segment in each group:

- Group G3 (control): t = 24h in saline solution,
- Group G4: t = 24h, after immersion in stain solution at concentration c3 = 15 g.L-1,
- Group G5: t = 24h, after immersion in stain solution at concentration c6 = 30 g.L-1.

Mann-Whitney tests were performed to assess the statistical validity of the (non-)effect of the agent on the mechanical properties, by comparison of G4 and G5 to the control group G3.

|  | Number of samples | Groups, names, and samples per group | | Remarks |
|---|---|---|---|---|
| Study of stain penetration | 24 | saline solution | 2 | immersion time 0h, 24h, 48h |
|  |  | c1 (5 g.L$^{-1}$ SPT) | 6 | immersion time 0h, 24h, 48h |
|  |  | c2 (10 g.L$^{-1}$ SPT) | 6 | immersion time 0h, 24h, 48h |
|  |  | c3 (15 g.L$^{-1}$ SPT) | 6 | immersion time 0h, 24h, 48h |
|  |  | c4 (20 g.L$^{-1}$ SPT) | 2 | immersion time 0h, 24h |
|  |  | c6 (30 g.L$^{-1}$ SPT) | 2 | immersion time 0h, 24h |
| Study of mechanical effects of contrast stain | 18 | G3 (control) | 6 | saline, immersion 24h |
|  |  | G4 | 6 | c3 = 15g.L$^{-1}$, immersion 24h |
|  |  | G5 | 6 | c6 = 30g.L$^{-1}$, immersion 24h |
| *In situ* tensile tests | 6 |  | 3 | c3 = 15g.L$^{-1}$, immersion 24h |
|  |  |  | 3 | other non-optimal staining conditions (preliminary feasibility tests) |

Table 1: Summary of sample characteristics for the entire study (penetration, mechanical effect of the contrast agent, *in situ* tensile tests).

## 3. Results

### 3.1 Protocol adjustment

Figure 2 shows representative (cross-sectional) images obtained in XRCT from which cross-sectional area and contrast agent penetration were quantified. The mean total cross-sectional area increased with immersion time for all solutions used (saline, c2, c3), and Wilcoxon tests showed that this swelling effect was statistically not different between any stain solution and the saline solution ($p > 0.1$).

Figure 3 shows the increase of the stained area ratio with immersion time. The percentage of the stained cross-sectional area increased with immersion time and with agent concentration (46.7 %, 64.9 % 76.0 % respectively for c1, c2 and c3 at 24 hours and 62.8 %, 79.3 %, 89.7 % at 48 hours). Significant differences ($p < 0.036$) were found for each concentration, and for each time of penetration (p < 0.036), except between c2 and c3 after 24 hours of penetration (p = 0.063). No difference was found in stained depth in tangent or radial directions (p > 0.25).

Note, also, that it was observed that musculo-elastic units were less clearly distinguishable after 48 hours in the contrast agent (cf. Figure 4).

The mechanical effect of the staining technique is addressed now. Curves of force versus displacement are plotted in Figure 5 for all tested samples. Note that two samples had to be removed from the analysis, being obvious outliers (probably due to collection too close to the abdominal aorta). Mann-Whitney tests showed no statistical differences between different stain solutions regarding maximum forces (p > 0.15), maximum displacement (p > 0.24) reached at failure. It was concluded that this contrast agent had no significant effect on the mechanical properties of interest here.

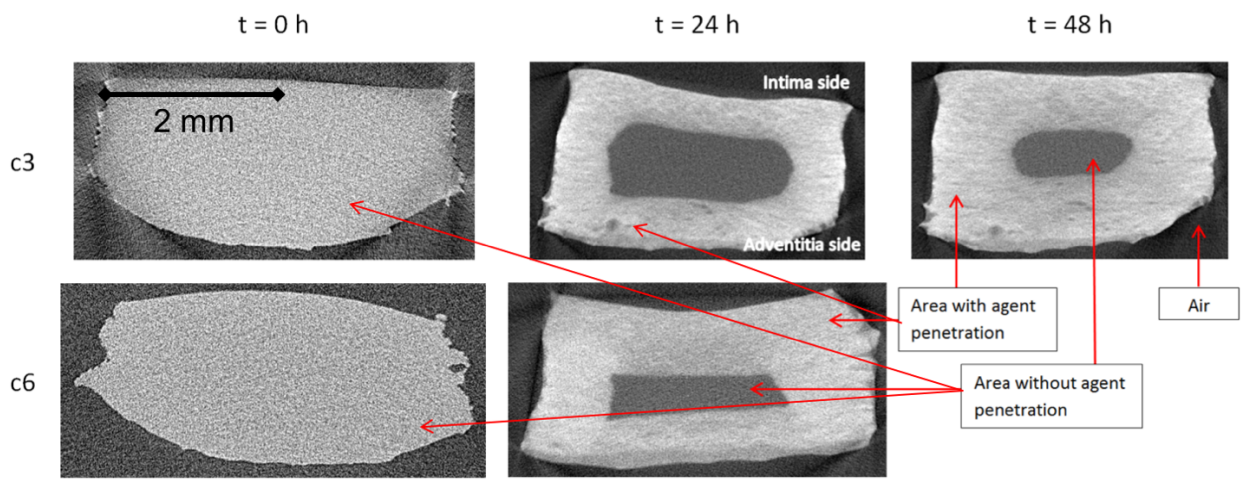

Figure 2: Typical XRCT images of longitudinal sample at t=0 h and after 24 and 48 hours in concentrated (c3 = 15 g.L$^{-1}$ and c6 = 30 g.L$^{-1}$) SPT. The resolution is 7 µm. The top of each imaged sample corresponds to the intimal side and the bottom to the adventitial side. The black area surrounding the sample is air and the inside dark area corresponds to the area contrast agent did not reach. Elastic sheets are visible in the bright area irrigated by the contrast agent. c6 at 48h was not performed because of the fuzziness already observed at 24h. Note that contrast was enhanced in images at t=0h, the actual X-ray absorption coefficient being the same as the inside dark area of other samples.

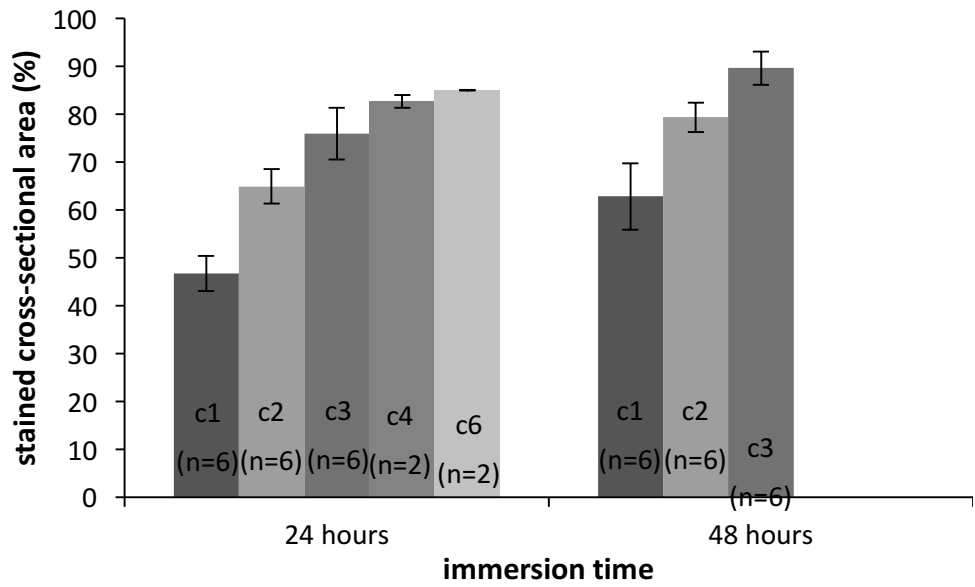
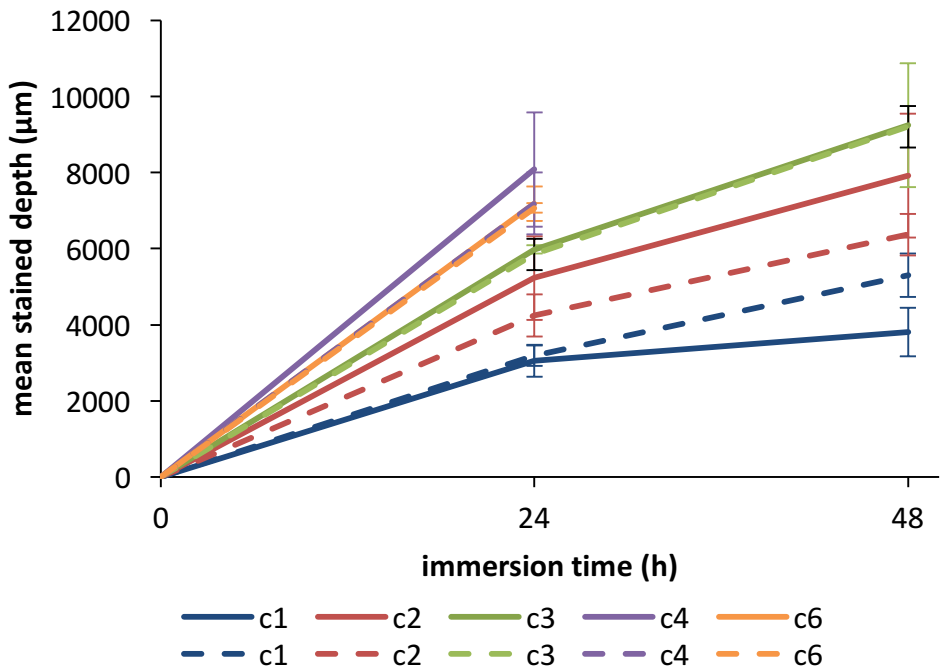

Figure 3: Study of contrast stain penetration. (top) percentage of stained cross-sectional area vs. immersion time. The error bars represent the standard deviation for each concentration. (bottom) mean stained depth vs. immersion time. The full lines represent the radial penetration, and the dotted lines the tangent penetration. n = 6 for each concentration.

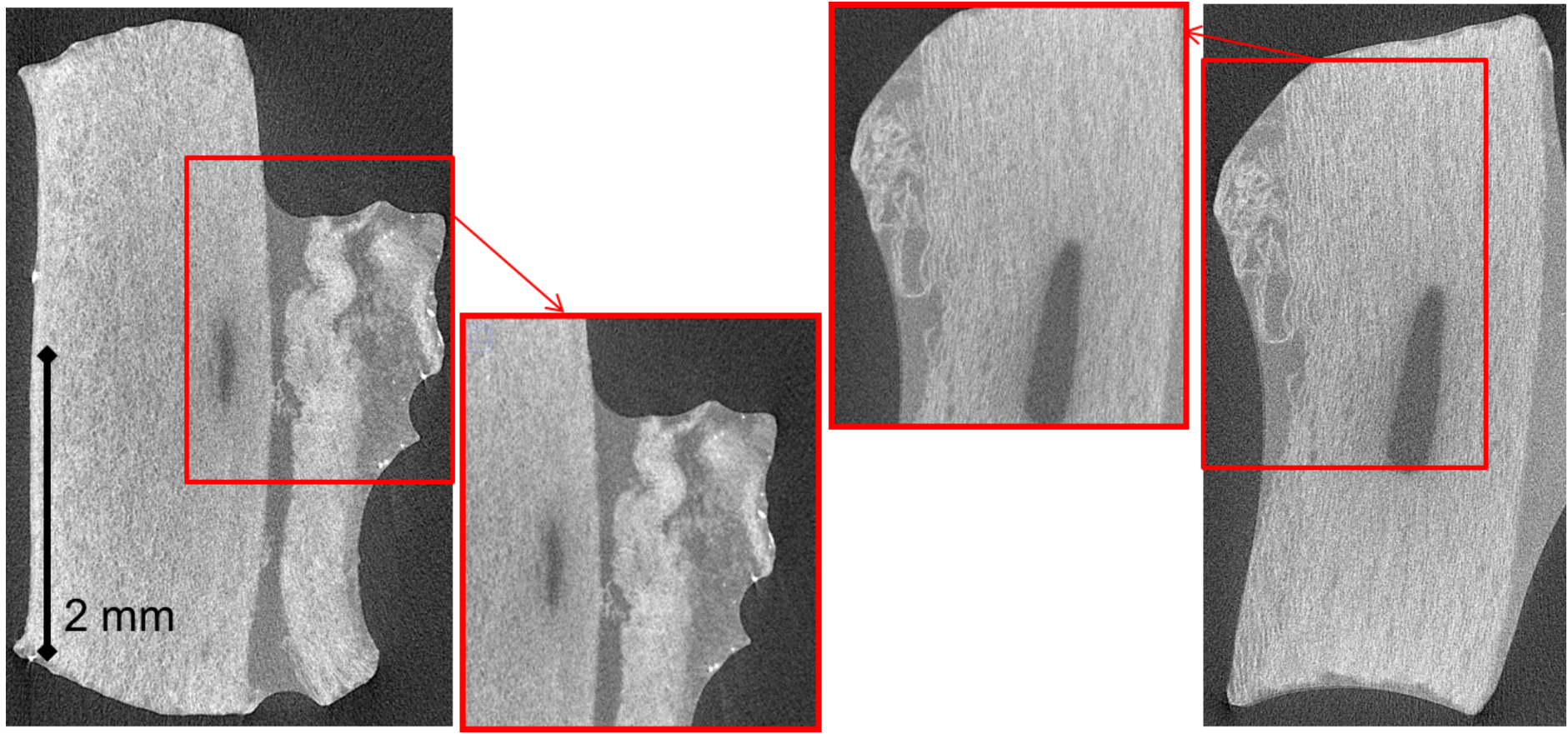

Figure 4: Visual comparison of the sharpness between 3D images obtained with samples immersed during 24 hours (right), and 48 hours (left) (4 µm voxel size).

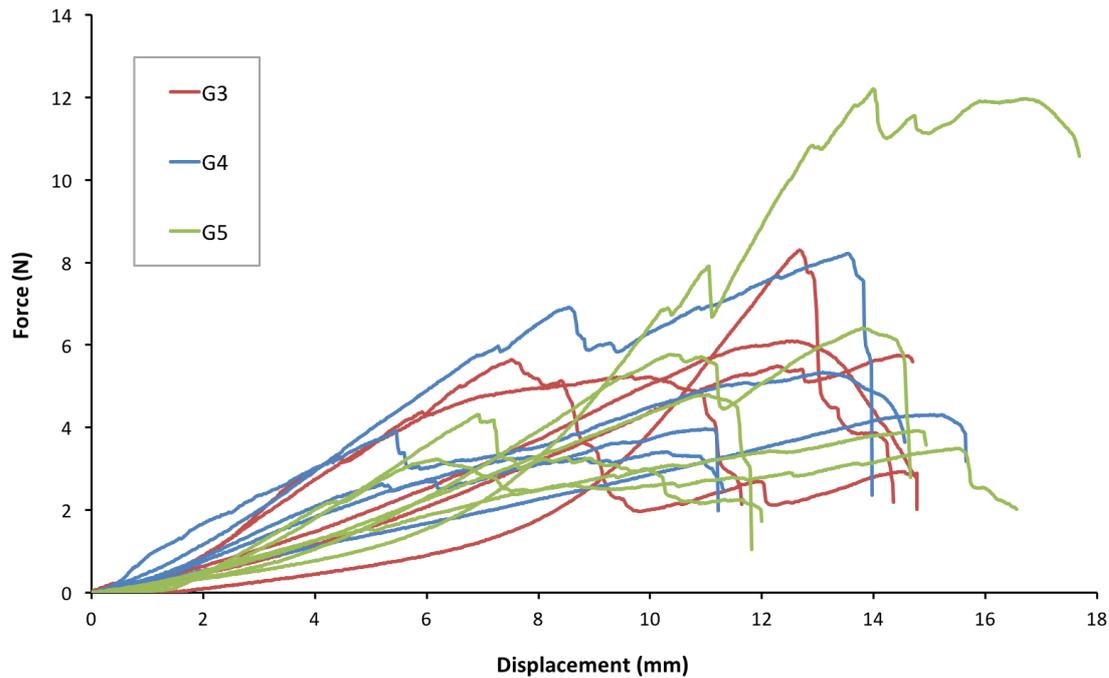

Figure 5: Mechanical effect of the contrast stain: force-displacement curves of all samples. G3 samples belong to the control group which was immersed in saline solution, G4 samples belong to the group which was immersed in 15g.L-1 SPT solution and G5 samples belong to the group which was immersed in 30g.L-1 SPT solution.

## 3.2 Observation of damage propagation

The experimental conditions and protocol which were adjusted and presented in the previous section were used to qualitatively describe the propagation of damage in porcine medial arterial tissue. For this purpose, three samples were tested *in situ* using the best compromise regarding the use of the contrast agent: 15g.L-1 concentration and 24 hour immersion. Note that three samples were also tested with non-optimal conditions of immersion (tests performed as a preliminary work of feasibility), these tests were also considered to draw the qualitative observations described below.

The observed damage mechanism was the same for all samples, whatever the sample orientation, the penetration time, or conservation solution (including the four pre-test samples). A sudden fracture of a group of musculo-elastic units perpendicular to the tension direction, was observed. This suggests a mode I rupture as a local initiation phenomenon. The elastic recoil of these ruptured units was accompanied by a progressive delamination of the tissue where this group of units separated from still intact units. This clearly indicates mode II separation as a preferred propagation of damage. This process was then repeated until all units were ruptured. Note, also, that this process initiated from the intimal side and then propagated, with successive rupture of units or groups of units. This phenomenon is clearly illustrated from the sequence of radiographs shown in Figure 6 and in the video (see supplementary materials). A schematic illustration of this process is also proposed in Figure 7.

Histological investigations confirmed these observations. As can be seen in Figure 8, evidence of long longitudinal delamination planes could be clearly distinguished, propagating perpendicularly to the main radial crack (see large arrows) along distances much larger than the thickness of the sample. In addition, histology showed that a high number a short longitudinal delamination planes were also formed during the process of damage and rupture (see small arrows).

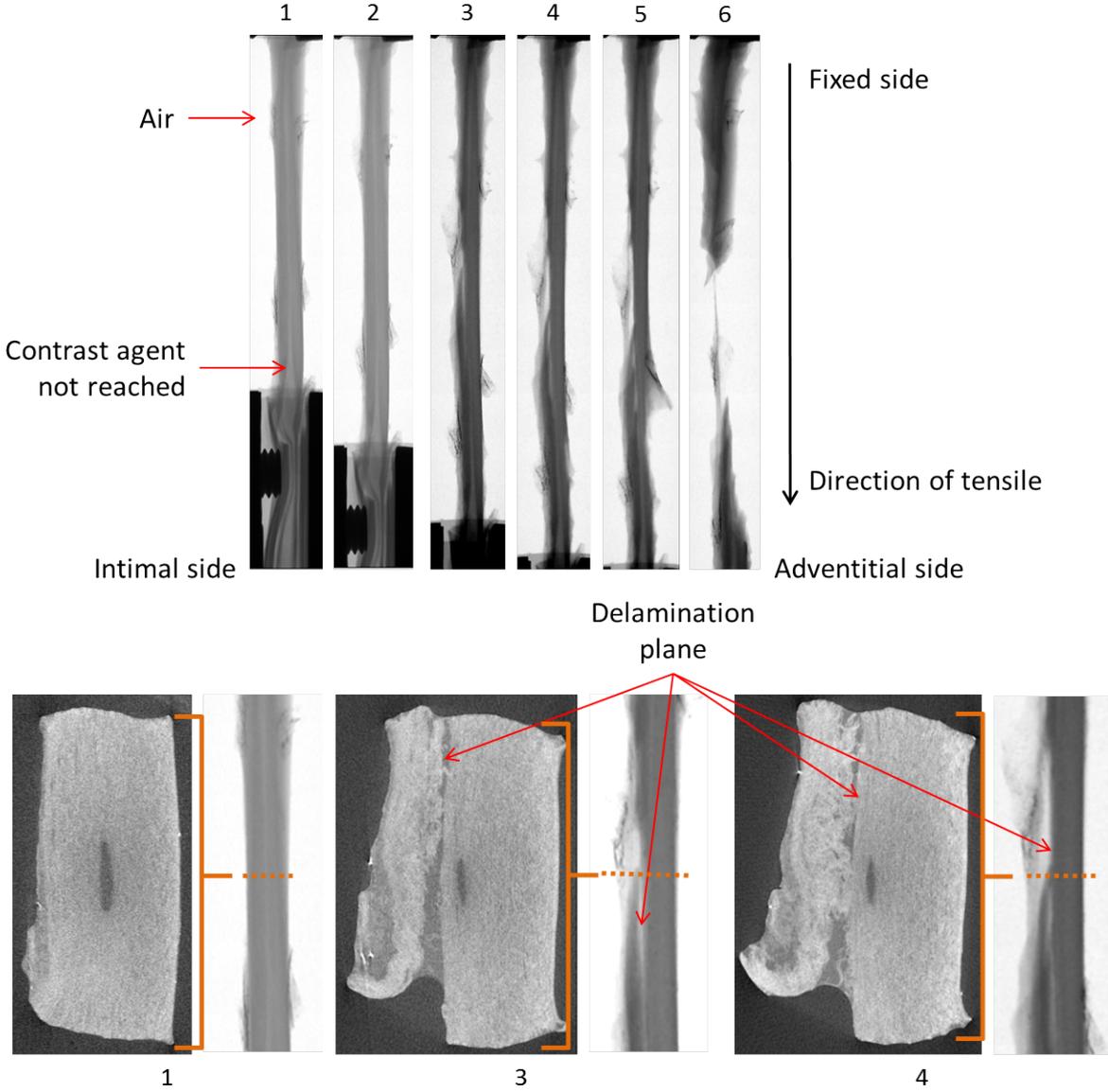

Figure 6: Rupture of a longitudinal sample with contrast agent (48 hours immersion, solution c3). (Top) successive radiographs illustrating the failure process, (bottom) cross-sectional views illustrating the microstructure at the initial state (1) and after rupture of the first elastic sheets (3-4). The location of the cross-section is indicated by dashed segments on the corresponding longitudinal views. Voxel size of 4 µm. Note that stain penetration was not complete for this sample.

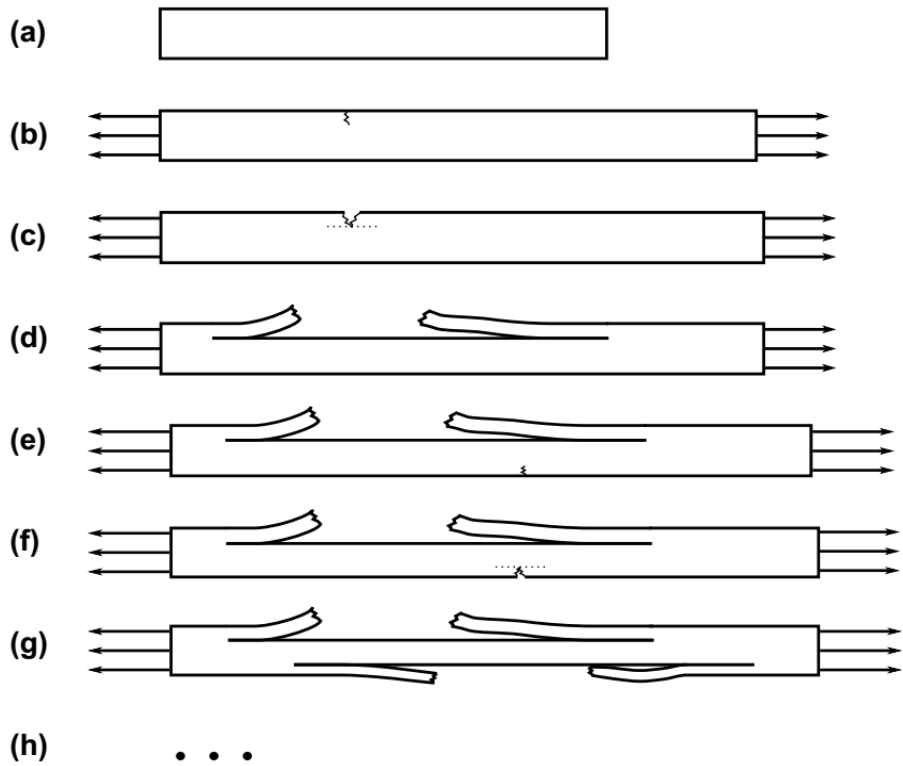

Figure 7: schematic representation of the damage initiation and propagation mechanism observed *in situ*. (a) intact sample, (b and c) initial radial crack, opening in mode I, (d) elastic recoil of the ruptured layers, causing a mode II longitudinal crack to form and propagate, (f, g, h...) the process repeats until complete failure of the sample.

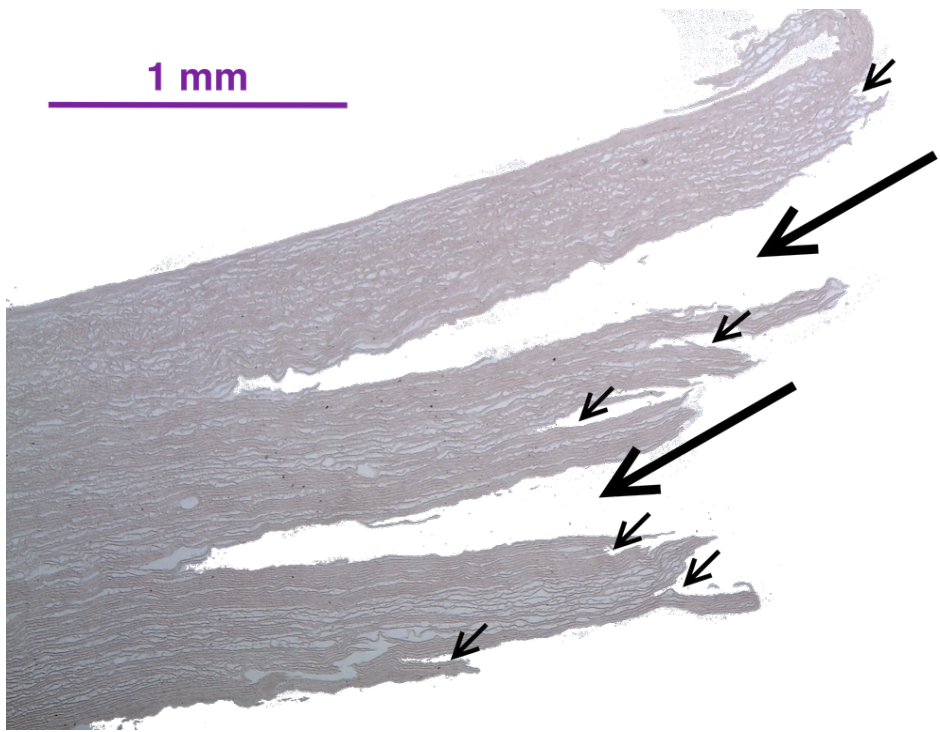

Figure 8: Histological section of a representative longitudinal cross-section in the region of the main crack of a sample. Large arrows show long delamination planes, short arrows show some short delamination planes.

# 5. Discussion

This paper presented a novel experimental approach to study structure-to-mechanics relationships in arterial soft tissue, focusing on the rupture of the medial layer. This approach is based on XRCT which is an accessible technique which revealed itself relevant for these tissues when used in combination with a contrast agent under suitable conditions that are detailed herein. The proposed conditions may constitute a basis for investigations on other types of soft tissues.

It is the first time, to the best of the authors' knowledge, that a detailed observation and description of the initiation, delamination and rupture process of medial tissue under tensile loading is described. This phenomenon, observed *in vitro*, is of particular interest from a clinical point of view. Indeed, a frequent and serious vascular disease is the aortic dissection. This pathology is initiated by a sudden and rapidly growing tear through the medial layer of the aortic wall. It is known that this event is mechanically triggered at the location of a local defect in the medial layer, and involves a progressive delamination of the medial musculo-elastic sheets, a tear in the intimal layer, consequently leaving a path for blood to flow through (Davies M.J., 1998; Sommer et al., 2008; Criado F.J., 2011). The present experimental approach may constitute a relevant model for the *in vitro* study of this specific rupture mechanism.

The following limitations to the present study should be mentioned. First the experiments were performed in non-controlled air. The testing chamber was small (only a few cubic centimeters) and contained soaked cotton pads to maintain humidity-saturated air and avoid drying the samples. As this process was not controlled, some discrepancies may have arisen due to partial drying, though it is not possible to quantify them. Second, the tests performed in this study were uniaxial. Biaxial tests, like inflation-extension tests on whole arterial segments, would allow reproducing loading conditions closer to the *in vitro* reality. This requires the development of a specific device to be used in the XRCT environment and constitutes an interesting perspective to this work.

Other interesting perspectives comprise, for instance, the use of different contrast agents to observe different components like collagen in the adventitial layer. This would permit to address rupture mechanisms which are specific, for instance, to aneurysms known to have collagen structural disorders. Also, the use of very high resolution equipment to enable more localized analysis of tissue deformation is an attractive perspective towards a better understanding of arterial micro-mechanics and cell mechano-sensing. In brief, this technique is very promising for further investigations in soft tissue structure to mechanics relationships.

# Acknowledgements

This work was supported by the European Research Council [starting grant n°638804, AArteMIS].